\documentclass{ws-p8-50x6-00}

\begin{document}

\title{Deconfinement and non--zero baryon density}

\author{Maria-Paola Lombardo}

\address{ Istituto Nazionale di Fisica Nucleare \\ 
Laboratori Nazionali del Gran Sasso \\I-67010 Assergi (AQ), Italy\\
lombardo@lngs.infn.it}

\maketitle

\abstracts{I discuss a few issues related with deconfinement  at finite baryon 
density by considering  lattice results for two colors
QCD, and ``toy'' studies of three colors QCD.} 

Substantial progress has been achieved 
in the understanding of the phase diagram of models that 
share the same global symmetries of QCD \cite{Phase}. 
Confinement, however, is not handled in
a completely satisfactory way by these studies.
Among the approaches  which include confinement, the 
monomer--dimer--polymer representation
of the partition function \cite{mondim} is limited to  the infinite
gauge coupling limit, where the dynamics is far from realistic, 
and the nature of the phase transition
is completely dominated by lattice artifacts. 
Polyakov loop models \cite{Olaf}, on the other hand, 
recover the quenched limit
of QCD and incorporate confinement also in the continuum.
Unfortunately, as these studies rely on large quark masses  
they cannot describe the chiral transition. Approaches
based on the Dyson-Schwinger equation, which in principle can
treat both chiral symmetry breaking and confinement, are intrinsically
approximate\cite{rs}.

Summing up (and oversimplifying), at finite
density chiral symmetry and mechanisms of confimenent
have been studied independently,  but little is known about their 
interplay, at least not from first principles. Such
interplay  lie
at the very core of the physics of the problem
-- after all,
deconfinement and chiral symmetry restoration
at finite density are intimately related with
asymptotic freedom, and long distance screening.
In addition,  it has
also been suggested that the exact realisation of confinement
might be crucial for a successful algorithm \cite{KLS0}. 

A so far unique possibility to
study a gauge model at finite density is afforded by
two colors QCD, which can be studied via numerical simulations even
at non--zero chemical potential\cite{HKLM}. I shall first discuss the 
interplay of thermodynamics, deconfinement
and chiral symmetry in this model, then
I shall try to assess the chances
to address similar questions in real (three colors) QCD.

\section{Two colors}
 
In Fig. 1 I sketch a possible (i.e. consistent with the
numerical results  described below) phase diagram of two colors QCD,
in the chemical potential--temperature plane, for
an arbitrary, non--zero bare quark mass.

\begin{figure}[t]
\epsfxsize=20pc %
\epsfbox{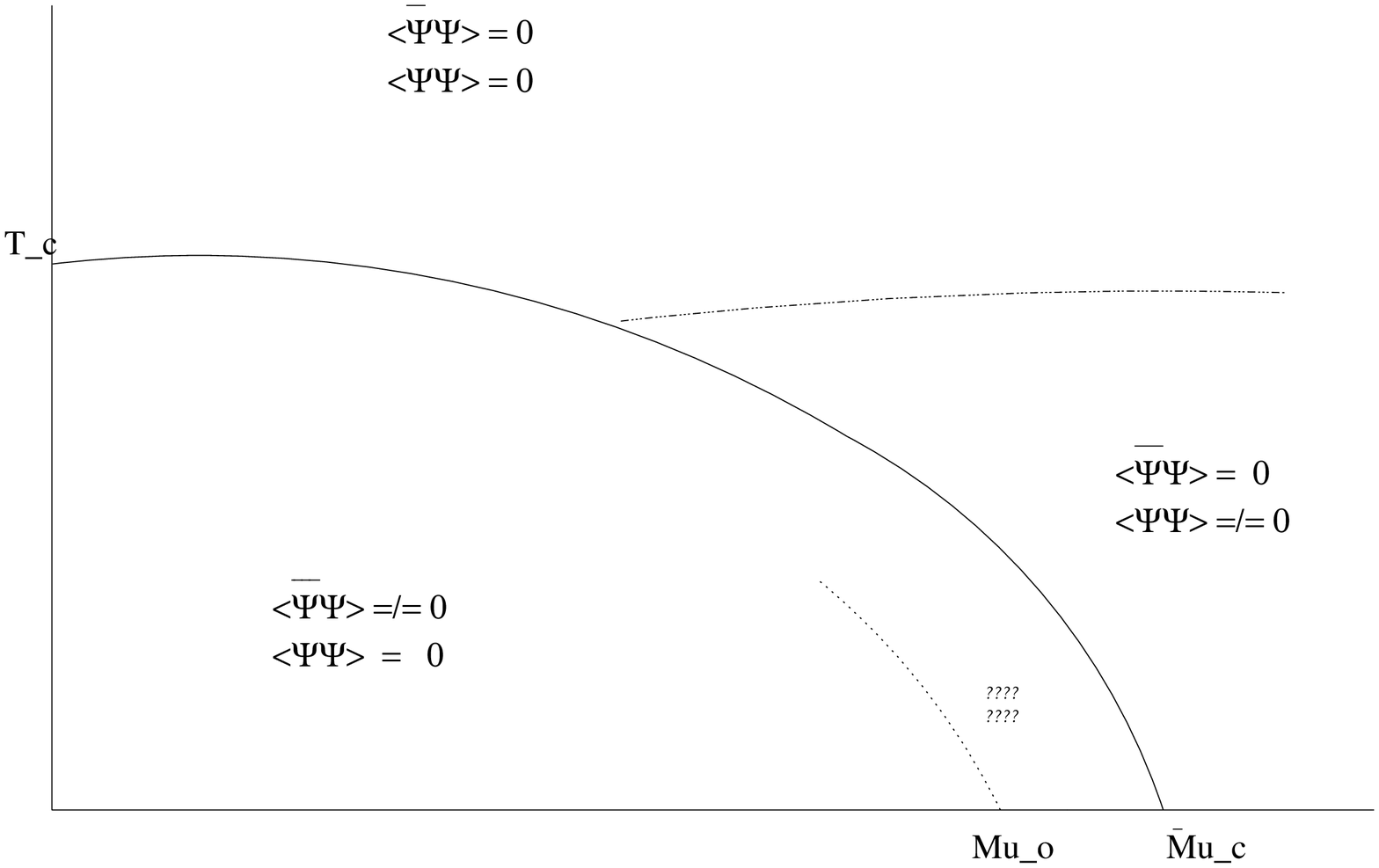} 
\caption{A sketchy view of the phase diagram of two colors QCD for
a non--zero quark mass}
\end{figure}

The $\mu=0$ axis has been investigated in \cite{john}, 
while  the $T \simeq 0.$ axis  
has been studied in \cite {HKLM}, and discussed  at this meeting 
\cite {sisu}. I stress that both axes, even the 
$\mu=0$ one \cite{su2varie},  present a number of 
interesting open problems. The study presented here, which
attempts at reconstructing the complete phase diagram, is even more 
exploratory. None the less, I feel that some interesting aspects are 
already emerging at this stage.
  




Let me start by considering what happens while increasing chemical
potential at fixed temperature. 
According to the standard scenario, at zero temperature,
for a chemical potential $\mu_o$
comparable with the mass of the lightest baryons, 
such baryons start to be produced 
thus originating a phase of cold, dense matter. 
For $SU(2)$ baryons (diquarks)  are bosons 
(as opposed to the fermionic  baryons of real
QCD). There are then important differences
between the dense phase in $SU(2)$ and $SU(3)$,
as, obviously, the thermodynamics of interacting
Bose and Fermi gases is different. In particular, diquarks
might well condense, however partial quark liberation is possible
as well.   We might then expect a
rather complicated ``mixed'' nuclear matter phase, perhaps
characterised by both types of condensates -- the one
marked with question marks in Fig. 1 (such mixed phases are
also predicted by more detailed instanton studies \cite{new}). 
Infact, on the cold lattice (Fig. 2, left)  a pure cubic term,
expected of a cold phase of massless free quarks,  does not describe 
the behaviour of the number density. 
We do not have pure Bose condensate either, thought.
For $\mu > \mu_c$
our results \cite{HKLM} 
suggest $ <\bar \psi \psi> = 0$, i.e. pure diquark condensation.
At the same time (see below) 
indications of long distance screening and deconfiment become
more pronounced, so the behaviour might get closer to a pure
free quark phase, which, however, does not seem supported by the data:
as the diquark states appear to be bound \cite{HKLM},
their condensates can still influence the thermodynamics. A direct
measure of diquark condensate should completely clarify this point
\cite{sisu}.
A ``pure'' free quark phase, 
with complete  restoration of chiral symmetry (i.e. $<\bar \psi \psi>
= <\psi \psi> = 0$) could be reached  at even 
larger $\mu$ --  as this region is dominated by lattice saturation artifacts,  
improved/perfect actions \cite{wuwe} might be necessary to explore it.

Despite these uncertainties, it seems anyway clear 
that  screening and deconfinement compete against
condensation, and this is better seen on  on a ``warmer'' lattice, close
to $T_c$ : here, 
$n \propto \mu^3$, consistent with a free, massless quark gas
(with a somewhat surprisingly
small temperature contribution \cite{dirk}), 
suggesting the existence of a critical temperature for diquark 
condensation (i.e. a temperature beyond which diquarks will not condense
at any value of the chemical potential) smaller than $T_c$.


To obtain a more direct probe of deconfinement, we can look at
the interquark potential by calculating the correlations 
$ <P(O) P^\dagger (z)>$ of
the zero momentum Polyakov loops, averaged over spatial directions.
This quantity is related to the string tension $\sigma$ via
$ <P(O) P^\dagger(z)> \propto e^{-\sigma z}$.


We show the results for the Polyakov loop
correlators in Fig \ref{fig:polcor}, where we compare the
behaviour at various temperature  with that at various
chemical potentials. In both cases 
the trend suggests increased fermion
screening, string breaking  and the passage to a deconfined phase. 


\begin{figure}
\epsfxsize=15pc %
\epsfbox{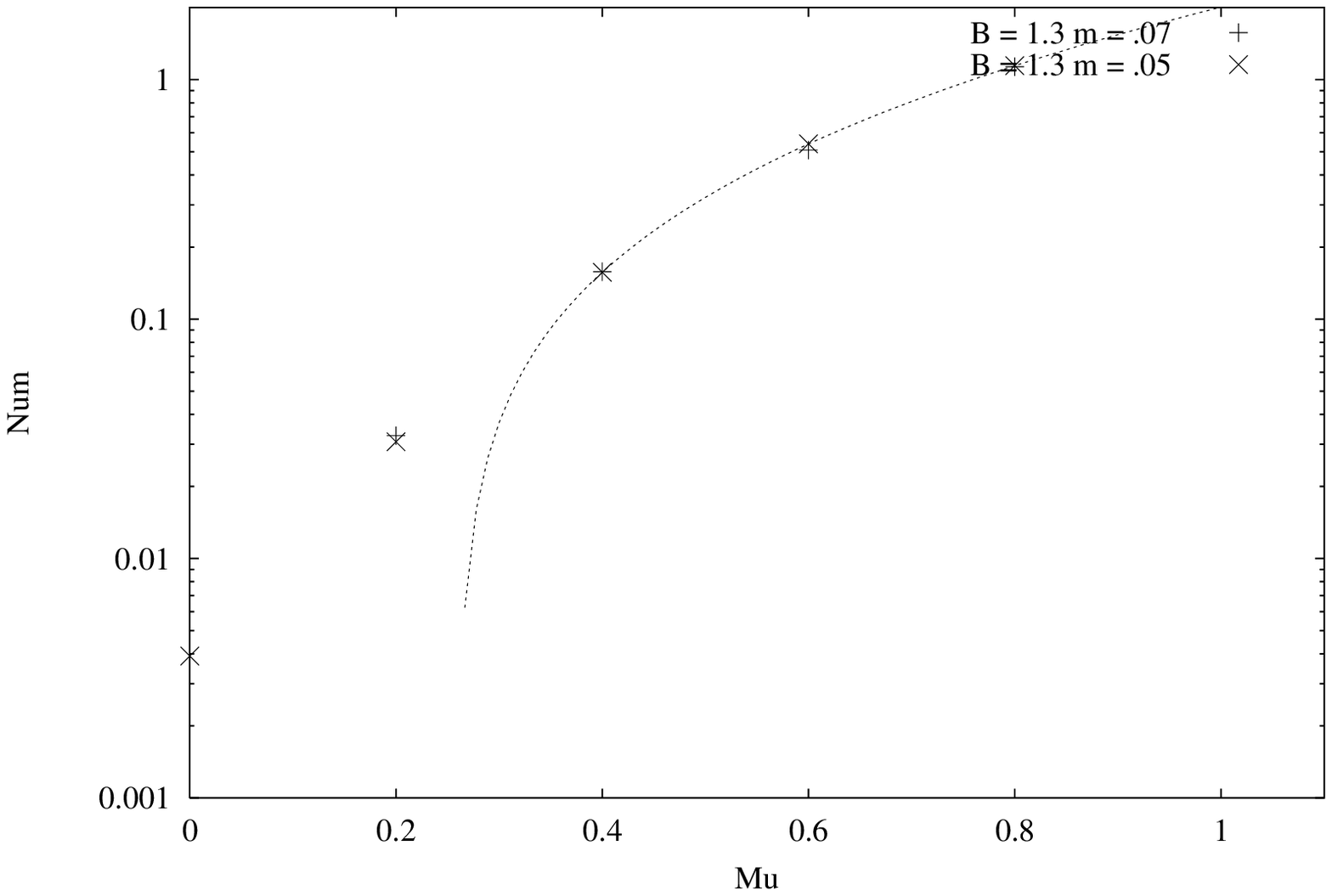} 
\epsfxsize=15pc %
\epsfbox{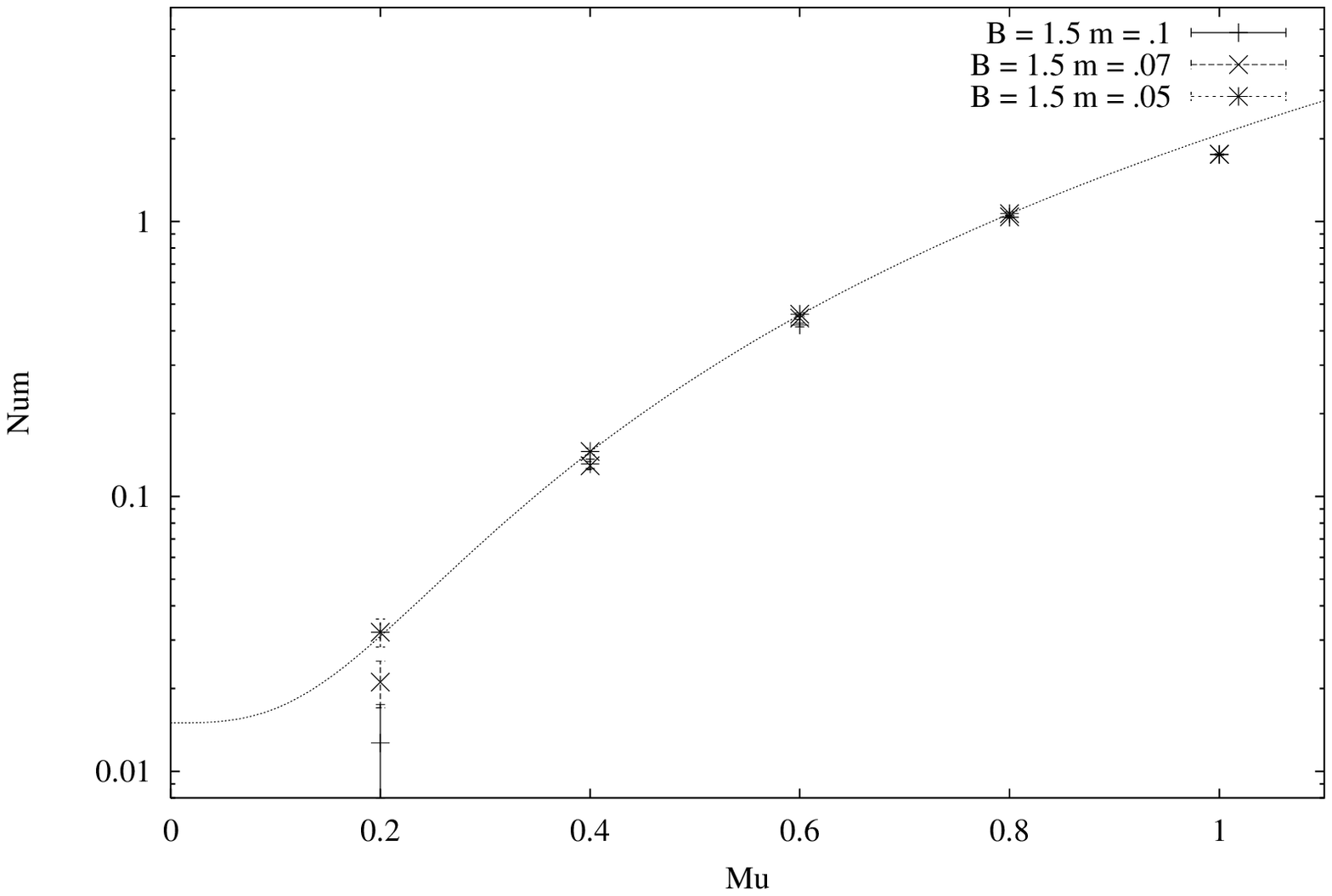}
    \caption[xxx]{Number density as a function of
the chemical potential, for different masses, on a cold (warm)
lattice on the left(right) hand side.
The cubic fits to $a\mu^3 + b\mu^2 +c$ are superimposed. 
b, c $\simeq 0$  on the warm lattice, consistent with a pure 
massless quark gas, while a `mixed' phase is possible on the cold lattice
(see text)}
\label{fig:lognum_vs_mu}
\end{figure}%
\begin{figure}[t]
\epsfxsize=15pc %
\epsfbox{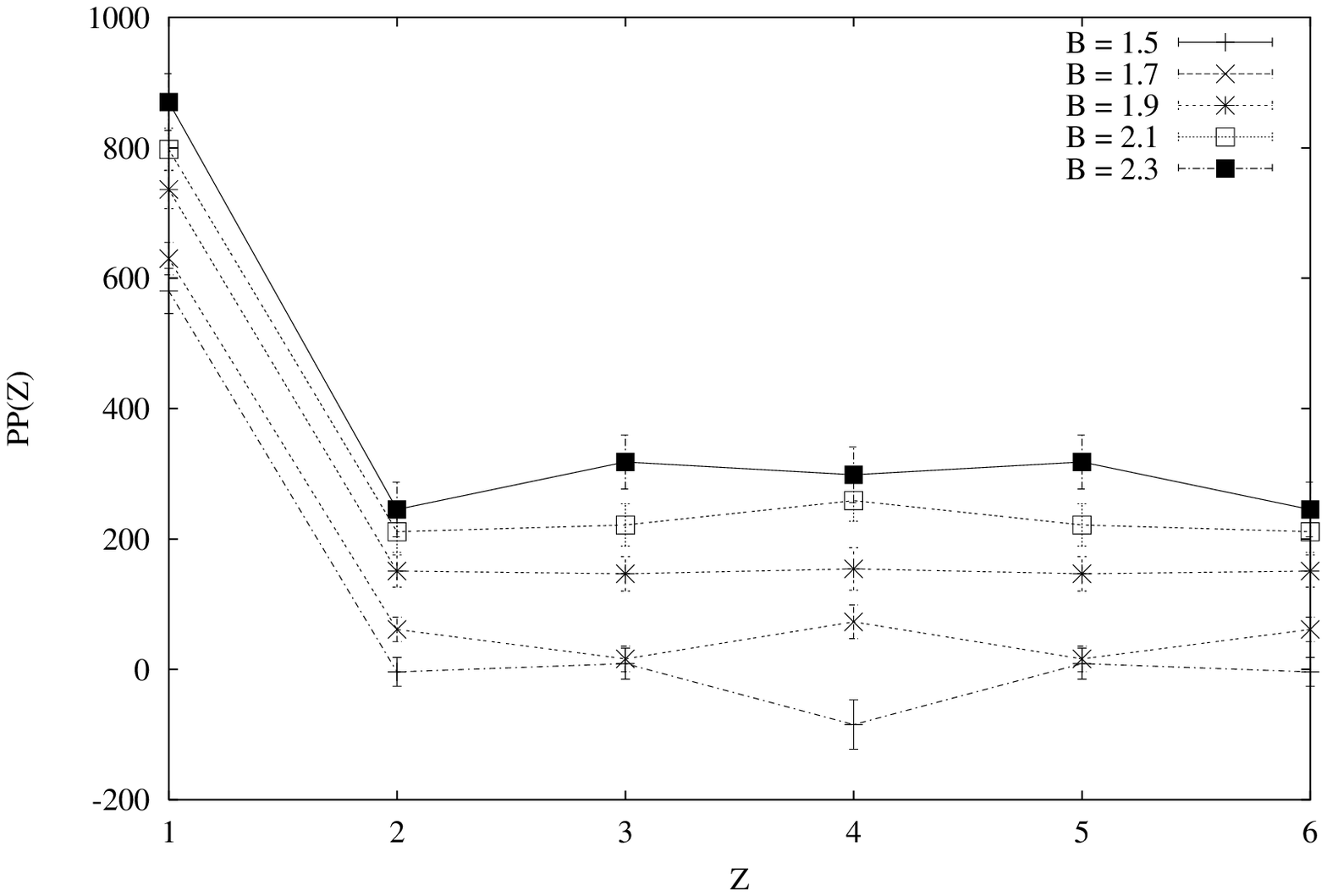} 
\epsfxsize=15pc %
\epsfbox{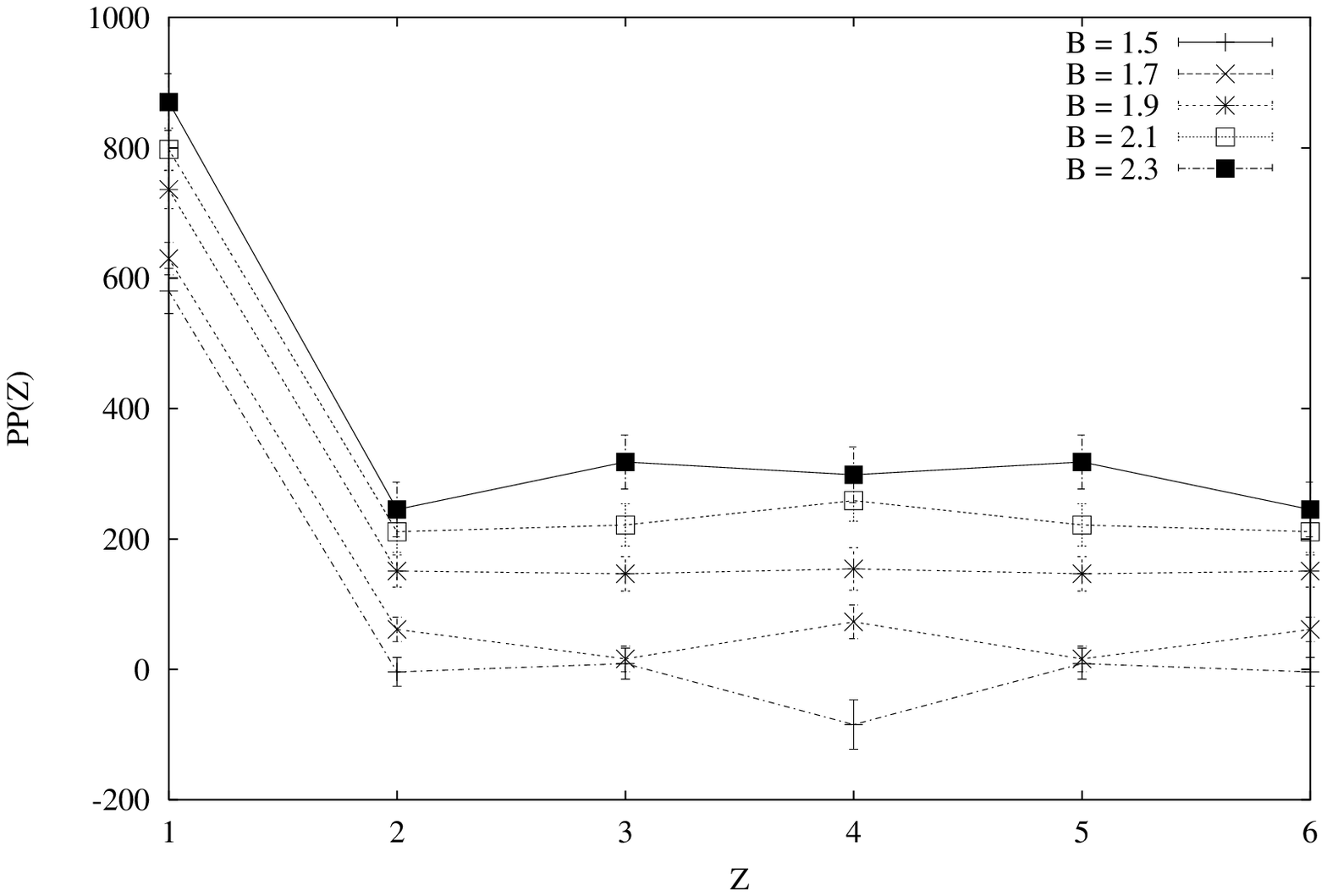}
\caption[xxx]{Correlations of the zero momentum Polyakov loop as
a function of the space separation. The left diagram is for
$\mu = 0$, and $\beta$ as indicated. The righthand part is for $\beta =
1.5$ and $\mu$ as indicated. In both cases we observe long
range ordering possibly associated with deconfinement}
\label{fig:polcor}
\end{figure}






\section{Two and three colors}

%
To gain further insight into the ``dynamical''role of the chemical
potential, and its effect on the gauge fields, we can take a look at a 
Toy \cite {BHKLM} version of the model, obtained by
inverting the Dirac operator at nonzero chemical potential in a
background of gauge fields generated at zero chemical potential.
For these configurations the potential is always 
the same as at $\mu = 0.0$. Interestingly, we have found that in this case
the behaviour of the diquarks propagators 
resembles that of the infamous quenched $SU(3)$
``baryonic pions''  measured in \cite{KLS}. 
This confirms that the nature of the interquark forces and
the deconfinement transition might well play
a major role  in $SU(2)$ and $SU(3)$ alike.
It is essential to
have the correct quark--quark
and quark--antiquark forces, 
since they  control and soften diquark condensations, including the
pathological ones. The inclusion of the 
chemical potential into the dynamics seems mandatory.

This brings us back to the necessity of 
first principle calculations
of QCD at finite density, hence to 
the problems with complex actions.

\section{Three colors}

Attempts at beating such problems fall in two main categories. 
Firstly, simulations at $\mu=0$
\cite{ian}, or imaginary chemical potential (which does
not systematically bias the ensemble)\cite{AKW}, combined 
with reweighting and/or analytic continuation. 
The main problem encountered here is that
the ensemble does not overlap with the non--zero density state of interest.
 Because of this, {\em physical} transitions might
{\em disappear}. Other approaches (as in the quenched approximation, or
calculations which use the modulos of the 
determinant \cite{mod}) 
include conjugate quarks so to keep the action
real when the chemical potential is included. 
The main problem here is the generation of
light particles with baryonic number (baryonic pions). Because of this,
{\em unphysical} transition might {\em appear}. 

There are reasons to believe that both problems might be alleviated
in the proximity of the $\mu=0$ deconfinement phase transition. 
Firstly, baryons will become
lighter, then easier to fluctuate. Secondly, baryonic pions will become 
heavier, and possibly decouple or dissolve
(clearly, one will have
to pay attention to the counting of degrees of freedom).

Consider the partition function in the $g=\infty$ limit of QCD \cite{strong}:
$$Z(\lambda. \mu) = 
2cosh(N_tN_c\mu) + sinh[(N_t+1)N_c \lambda] /sinh(N_t \lambda)$$
where the variational parameter $\lambda$  (essentially, $<\bar \psi \psi>$) 
is to be determined
by a minimum condition, $N_t$ is the number of points in time direction,
$N_c$ the number of colors. When $N_t$ grows large 
$ Z \simeq  e^{N_t N_c \mu} + e^{N_t N_c \lambda}$ : 
the critical chemical potential is a measure of
the strength of symmetry breaking.

For a purely imaginary chemical potential, 
$cosh(N_t N_c \mu) \rightarrow cos(N_t N_c\mu)$. 
We see the expected periodicity $ 2 \pi /(N_t N_c)$,
and we note that the chemical potential term can be ignored
for large $N_t$ (zero temperature). Indeed, we have verified that in
this limit any dependence on the chemical potential is lost 
\cite{fl} : clearly, 
$<\bar \psi \psi>$ as a function of (complex) $\mu$ is a constant in the
half plane $\Re(\mu)  < \mu_c$, $\mu_c$ being a real number.
By increasing the temperature, the effective potential changes with
imaginary $\mu$, and the chiral condensate {\em increases}  
with imaginary chemical potential, as it should. 
However, for this effect to
be appreciable, one needs to be very close to the critical point
as the reader can  easily check by exploiting the above formula.

Can we guess where we should work in real QCD for imaginary $\mu$ 
to be useful? Clearly, we need a region with large $\mu = 0$ derivatives.
The fluctuations of baryons are measured by 
the baryon number susceptibility \cite {GLRST}
$$\chi(T,\mu) = \partial \rho (\mu, T) / \partial \mu
= \partial^2 log Z (\mu, T) / \partial ^2 \mu$$
Lattice results  indicate the range where 
$\chi(T,\mu=0)$ is significantly different from zero. This is 
the candidate region for performing imaginary $\mu$  
calculation below
$T_c$ -- a narrow, but not minuscule interval. Perhaps it is
also worth considering  that , as
$log Z (T, \mu) = K + \chi(T,0) \mu^2 + O(\mu^4)$, 
the baryon number susceptibility obtained by  numerical
differentiation of  the number density with imaginary
chemical potential should be the negative of the one measured
with standard methods. It should then be possible to assess
and use imaginary chemical potential calculations without use of
the Laplace transform, or other analytic continuation techniques.
This should open the possibility to study the effect of
a small chemical potential on the deconfinement transition and
to address issues such as spontaneous parity violation near 
$T_c$ \cite{dima}.

In the same ``hot'' region reweighting too might prove useful. To check that,
we have tested the Glasgow method 
in one dimensional QCD\cite{exact}, a solvable 
model without SSB, but with baryons, whose partition function  
is easily related to that of four dimensional QCD. At large ``temperature''
the Glasgow method reproduces the exact results \cite{touka},
thus supporting the idea that reweighting
methods can be successfully used in that regime. It should be noticed,
however, that an analysis of the Lee Yang zeros  shows that in this
case there is no pathological  onset \cite{patho}:
the Glasgow method only rearranges
zeros on circular patterns. It is reasonable to expect that the same
happens  in high temperature, four dimensional QCD.

The simple discussions presented here support qualitative 
arguments suggesting that imaginary chemical potential and reweighting
might be successful at large temperature, close to the deconfining
transition. However, they also confirm the feeling
that the existing algorithms will not be adequate to study the
low temperature, finite density regime
no matter how big the statistics is. Perturbative
approaches to QCD which use composite nucleons as fundamental
variables \cite{pa} 
might well offer a promising avenue for 
finite density calculations in the low temperature, confined phase.

\section*{Acknowledgments}

I would like to thank the organizers of a very interesting and pleasant
meeting and many of the participants, in particular Kurt
Langfeld, Harald Markum, Misha Stephanov
and Jac Verbaarschot,  for useful discussions.  
I also wish to thank  Ian Barbour, Philippe de Forcrand, 
Simon Hands, John Kogut, Susan Morrison and Don Sinclair for
an enjoyable collaboration, and Venia Berezinsky, John Kogut and 
Edward Shuryak for  criticism and suggestions.
This work was partly supported by 
the  TMR network {\em Finite Temperature Phase Transitions 
in Particle Physics}, EU contract no. ERBFMRXCT97-0122.

\end{document}